\documentclass[aps,prl,twocolumn,groupedaddress]{revtex4} 
 
\usepackage{graphicx} 
\begin{document} 
\title{Griffiths phases in the strongly disordered Kondo necklace model} 
\author{Tatiana G. \surname{Rappoport}} 
\author{Beatriz \surname{Boechat}} 
\author{Mucio A. \surname{Continentino} }
\email{mucio@if.uff.br} 
\affiliation{Departamento de F\'{\i}sica - Universidade Federal Fluminense\\ 
Av. Litor\^anea s/n,  Niter\'oi, 24210-340, RJ - Brazil} 
\author{Andreia \surname{Saguia}} 
\affiliation{Centro Brasileiro de Pesquisas F\'{\i}sicas \\ 
Rua Dr. Xavier Sigaud 150 - Urca,  Rio de Janeiro, 22290-180, RJ - Brazil.}

\date{\today}

\begin{abstract} 
The effect of strong disorder on the one-dimensional Kondo necklace model is 
studied using a perturbative real-space renormalization group 
approach which becomes asymptotically exact in the low energy limit. The 
phase diagram of the model presents a random quantum critical point 
separating two phases; the {\em random singlet phase} of a quantum disordered $XY$ 
chain and the random Kondo phase. We also consider an anisotropic version of 
the model and show that it maps on the strongly disordered 
transverse Ising model. The present results provide a rigorous microscopic 
basis for non-Fermi liquid behavior in disordered heavy fermions due to 
Griffiths phases. 
 
\end{abstract} 

\maketitle 
 
An understanding of the effects of randomness on the quantum critical point 
($QCP$) of the $d=1$ Kondo necklace ($KN$) model \cite{doni} is relevant for the 
study of disordered heavy fermions systems with non-Fermi liquid behavior 
\cite{miranda,castroneto,aronson}. Recently a {\em non-perturbative} real 
space renormalization group ($RG$) was presented showing that {\em weak disorder is an irrelevant 
perturbation} near the $QCP$ of a $d=1$, anisotropic, pure $KN$ 
model  \cite{tati1}. This result is in agreement with the 
generalized Harris criterion \cite{Harris,chayes} for irrelevance of disorder, $\nu >2/d$, where $%
\nu =2.24$ is the value obtained for the correlation length exponent of the 
$QCP$ of the pure anisotropic system \cite{tati1}. On the other hand different 
approaches have been proposed to describe the non-Fermi liquid behavior of 
disordered heavy fermions that rely on the {\em relevance} of disorder \cite 
{miranda,castroneto}. In order to settle this important point we investigate 
here the one-dimensional $KN$ model in the case of {\em strong disorder} 
using a generalization of a perturbative real space $RG$ approach. This is the 
Ma-Dasgupta-Hu method \cite{ma} which has been extended by 
Fisher \cite{fisher1,fisher2} and others \cite{qc} mostly for the study of random 
quantum spin chains and the random transverse-field Ising model ($RTIM$). In 
this Letter we obtain an exact mapping of the strongly disordered $KN$ model 
into a problem of quantum spin chains. 
For the anisotropic $KN$ model, the mapping is in the random transverse-field 
Ising model ($RTIM$). The general mapping allows to apply directly to the $KN$ 
problem the results of extensive work done recently on random quantum spin 
chains \cite{fisher1,fisher2,qc,iglu1,iglu2,iglu3, young}. This includes the 
existence of Griffiths phases with the characteristic singular behavior of 
different thermodynamic quantities at low temperatures. The present work 
provides a rigorous microscopic justification for the non-Fermi liquid 
behavior of disordered heavy fermions due to the existence of a Griffiths 
phase in the neighborhood of a random $QCP$  \cite 
{castroneto}. 
 
The one-dimensional $KN$ model is defined by the 
Hamiltonian, 
\begin{equation} 
H=\sum_{i=1}^{L-1}W_{i}({{\sigma^{x}}}_{i}{{\sigma^{x}}}_{i+1}+{{\sigma^{y}}}%
_{i}{{\sigma^{y}}}_{i+1})+\sum_{i=1}^{L-1}J_{i}{\vec{S}}_{i}.{\vec{\sigma}}%
_{i},  \label{hamil} 
\end{equation} 
where $\sigma^{\mu}$ and $S^{\mu}$, $\mu=x,y,z$ are spin-1/2 Pauli matrices 
denoting the conduction electrons and the spins of the local moments, 
respectively. The sites $i$ and $i+1$ are nearest-neighbors on a chain of $L$ 
sites. The local Kondo interactions, $J_{i}>0$ and the hopping energies $%
W_{i}>0$ are uncorrelated quenched random variables with probability 
distributions, $P_{J}(J_{i})$ and $P_{W}(W_{i})$. In the anisotropic version 
of the model, ($X-KN$), the band of conduction electrons is represented just 
by an Ising term, $\sum_{i=1}^{L-1} W_{i}{{\sigma^{x}}}_{i}{{\sigma^{x}}}%
_{i+1}$. The full isotropic $KN$ model in Eq.~ (\ref{hamil}) will be refered from now on as the 
$XY-KN$ to avoid confusion. 
 
The $KN$ model was proposed by Doniach \cite{doni} to study heavy fermions 
and emphasizes magnetic degrees of freedom neglecting charge fluctuations. 
It incorporates the essential physics of these systems which results from 
the competition between Kondo effect and magnetic ordering. In the absence 
of disorder, the models above have distinct behavior. For the $X-KN$ there  
is an unstable fixed point at a finite value of $(J/W)$ 
separating an antiferromagnetic phase from a spin compensated, Kondo-like 
phase  \cite{tati1}. For the $XY-KN$, 
any interaction $J>0$ gives rise to a dense Kondo  
state \cite{tati1,scalettar85,santini92,moukouri95,zhang00}. 
 
In this Letter, in order to implement the perturbative $RG$ method for the 
$KN$ models, we consider the conduction electrons $\sigma_i$ and the local 
moments $S_i$ arranged in a chain as shown in Fig.1. Next we choose the 
largest interaction in the chain, 
\begin{equation} 
\Omega = \max \{ W_i,J_i\} 
\end{equation} 
If the strongest interaction is a Kondo coupling ({\em bond}) between a 
local moment and a conduction electron, for example, $\Omega_I = J_2$, the 
local moment $S_2$ and the conduction electron $\sigma_2$ are decimated out 
yielding an effective hopping $\tilde W$ between the conduction electrons $\sigma _{1}$ 
and $\sigma _{3}$ at neighboring sites  which is obtained in 
second order perturbation theory (see Fig. 1). 
 
If the strongest interaction is a hopping term, say, $\Omega_I = W_2$, the 
four spins $\sigma_2$, $S_2$, $\sigma_3$ and $S_3$ are considered as a {\em %
cluster}. This is replaced by an effective two-spin cluster consisting of 
renormalized, local moment $\tilde S$ and conduction electron $\tilde \sigma$ 
coupled by a new effective local Kondo interaction, $\tilde J$ (see Fig. 1). Thus, after 
decimating a strong interaction, $W_i$ or $J_i$, we have an effective 
Hamiltonian with two less spin degrees of freedom and all couplings $< 
\Omega_I$. 
 
\begin{figure} 
 \includegraphics[width=8.5cm]{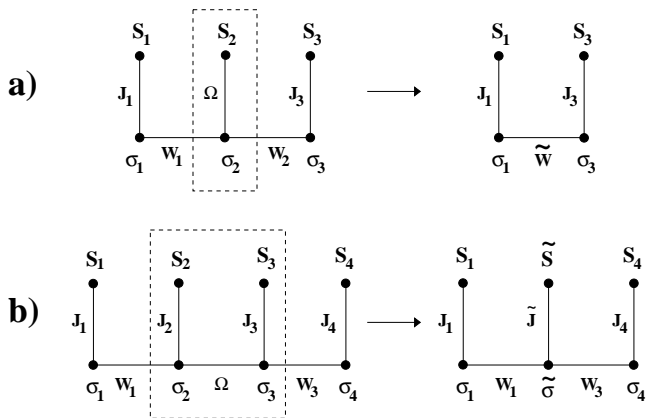} 
 \caption{\label{fig1}The decimation processes in the case the strongest 
interaction  $\Omega$ is either a) a bond  or b) a hopping .} 
\end{figure}

The $RG$ transformation gives, in the case the strongest interaction is a 
bond, an effective hopping, 
\begin{equation}  \label{potencia} 
\tilde W = {\frac{W_{1}~W_{2}}{\kappa \Omega}} 
\end{equation} 
and in the case it is a hopping we obtain, 
\begin{equation}  \label{hopping} 
\tilde J = {\frac{ J_2~J_3}{\kappa \Omega }} 
\end{equation} 
The new Hamiltonian has exactly the same form as the original one, but now 
the system is formed by spin clusters and effective bonds. Note that the 
resulting flow equations, Eqs.~(\ref{potencia}) and (\ref{hopping}), present a 
duality between $W$ and $J$.   We find that for the  $X-KN$ 
model the parameter  $\kappa=1$, so that the recursion relations for this 
model map exactly into those of the 
$RTIM$ \cite{fisher2}. For the $XY-KN$ 
model, we get $\kappa = 4/ \sqrt{6} \approx 1.63$. 
 
The method is implemented numerically on samples of sizes up to $L=2^{18}$ and 
averages over $10^2$ configurations. We use rectangular 
distributions for the local bonds and hopping terms. Periodic boundary 
conditions are applied. The relevant  parameter is the ratio $%
(J_{0}/W_{0})$ of the cut-offs of the original distributions. Furthermore we 
take $W_{0}=1$ such that $J_{0}$ is taken as the control parameter. The dual 
nature of the recursion relations allows to locate the random $QCP$ 
at $J_{0}=1$ for any $\kappa$. 
 
We measure the distance to the random $QCP$ by the variable 
\begin{equation} 
\delta = \frac{< \ln J > - < \ln W >}{ var ( \ln J) + var ( \ln W ) } 
\end{equation} 
where $<->$ means average over quenched disorder and $var(x)$ 
denotes the variance. Of course $\delta=0$ for $J_0=1$. 
 
At the $QCP$, $J_0=1$ or $\delta=0$, the parameter $\kappa \ge 1$ is 
irrelevant \cite{senthil} and we obtain for both models a behavior 
associated with a {\em random singlet phase} of the local moments. 
This is characterized by the 
fact that the fixed point distributions of bonds, $P_{J}(J_{i})$, and 
hoppings, $P_{W}(W_{i})$, have power-law forms, 
\begin{eqnarray}  \label{alfaj} 
P_{J}(J_{i}, \Omega) = \frac{\alpha_J}{\Omega} \left( \frac{\Omega}{J} 
\right )^{1-\alpha_J} \theta( \Omega -J)  \nonumber \\ 
\nonumber \\ 
P_{W}(W_{i}, \Omega) = \frac{\alpha_W}{\Omega} \left( \frac{\Omega}{W} 
\right )^{1-\alpha_W} \theta(\Omega - W) 
\label{dist} 
\end{eqnarray} 
with the exponents $\alpha_W$ and $\alpha_J$ depending on the cut-off 
$\Omega$. They are given by, 
\begin{equation} 
\alpha_W = \alpha_J = - \frac{1}{ \ln \Omega } 
\end{equation} 
The physical behavior of the thermodynamic quantities as a function of 
temperature $T$, which arises from such distributions of interactions is 
extensively described in the literature \cite{fisher2,qc,fisher}. It is 
given by power laws with logarithmic corrections. 
 
For the anisotropic $KN$ model, the $QCP$ at $%
J_0=1$ separates a disordered antiferromagnetic phase ($J_0<1$) from a 
dense Kondo compensated phase ($J_0>1$). The phase for $J_0 < 1$ of 
the $XY-KN$ model, where  $P_J(J_i)$ becomes negligeable under iteration,  is 
identified, from the limit $J_0 =0$, as a  random singlet phase.  In fact  
this limit corresponds to 
the strongly disordered $XY$ quantum chain which, as shown by Fisher \cite{fisher1}, 
exhibits a  random singlet phase. 
The nature of the disordered Kondo compensated phase for $J_0 >1$ is very similar in both $KN$ models 
and will be discussed in detail below. 
 
Above and below, but close to criticality, the recursion relations in Eqs. 
(\ref{potencia}) and (\ref{hopping}), give rise to Griffiths phases 
with a range $0 <|\delta |<\delta_G$. 
 In each side of the $QCP$ such phases are dominated by rare, very 
large clusters of the opposite phase. They are also characterized by power 
law behavior of the probabilities distributions, as in Eq. (\ref{dist}),  but in 
this case with exponents which depend on the distance $\delta$ to the 
$QCP$.  As an illustration, we show in Fig. 2 the exponent $\alpha_J$, 
in the disordered phase ($J_0>1$) as a function of the cut-off $\Omega$, as 
$\Omega$ is reduced under iteration, for different distances ($\delta< \delta_G$) of the random 
$QCP$. For small $\delta$, sufficiently close to the $QCP$, the finite 
intercept for $(-1/ \ln \Omega) \rightarrow 0$ is directly proportional to 
this distance $\delta$  \cite{fisher2}. 
 \begin{figure} 
 \includegraphics[width=7cm,angle=270]{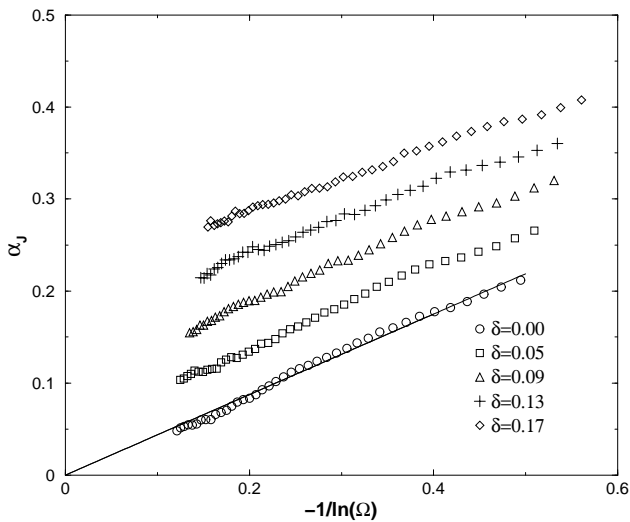} 
 \caption{\label{fig2} 
The exponents $\alpha _{J}$  of  Eq. (\ref{dist}) at the 
disordered dense Kondo phase ($J_{0}>1$) of the $X-KN$ model for different 
distances $\delta $ to the random $QCP$  in the case of 
rectangular gapless distributions. The abscissa is the inverse of the 
logarithm of the effective cut-off.} 
\end{figure} 
 
The range $\delta_G$ of the Griffiths phases in control parameter space depends on the 
original distributions of the interactions. For gapless distributions, i.e., 
those unbound from below,  the Griffiths phases extend all over the 
phase diagram \cite{iglu1,iglu2}. For bounded 
distributions from below, i.e., with a low energy gap, they 
have a finite extension  around the $QCP$. 
 
The divergence of the susceptibility in the Griffiths phase is 
given by, $\chi _{0}(T)\propto T^{1/z_{\kappa 
}(\delta )-1}$ where $z$ is the {\em dynamical exponent}  \cite{iglu1,iglu2}.  
The behavior 
of the specific heat in this phase is given by, $C_{V}(T)/T\propto T^{-1+1/z_{\kappa 
}(\delta )}$ and for small magnetic fields $H$, the magnetization, $M\propto 
H^{1/z_{\kappa }(\delta )}$. 
The $z$ exponent has the usual meaning of a dynamic exponent in quantum 
phase transitions in that it relates length and time scales. It is an invariant of the $RG$ 
equations, Eqs. (\ref{potencia}, \ref{hopping}),  and assumes the values 
$z= \infty$ and $z=1$ at the random $QCP$ ($\delta=0$) and at the border of the Griffiths  phase 
($|\delta|=\delta_G$), respectively\cite{iglu2,iglu3}. 
 
Fig. 3 shows the dynamic exponents $z_{\kappa }(\delta )$  for 
both $X-KN$ and $XY-KN$ models in the Griffiths phase, at  $J_{0}>1$,  for 
rectangular unbound distributions. Due to its invariance along the iteration process this 
exponent is very useful to characterize the temperature dependence 
of different thermodynamic quantities in all 
the Griffiths phase. For the original gapless distributions 
used to obtain $z_{\kappa }(\delta )$ in this figure, this phase extends 
for all $\delta >0$ \cite{iglu2,iglu3}.  Note from Fig.~3 that 
for sufficiently small $\delta $, $z_{\kappa }(\delta )^{-1}\propto \delta $ \cite{fisher1,iglu1}. 
\begin{figure} 
 \includegraphics[width=7cm,angle=270]{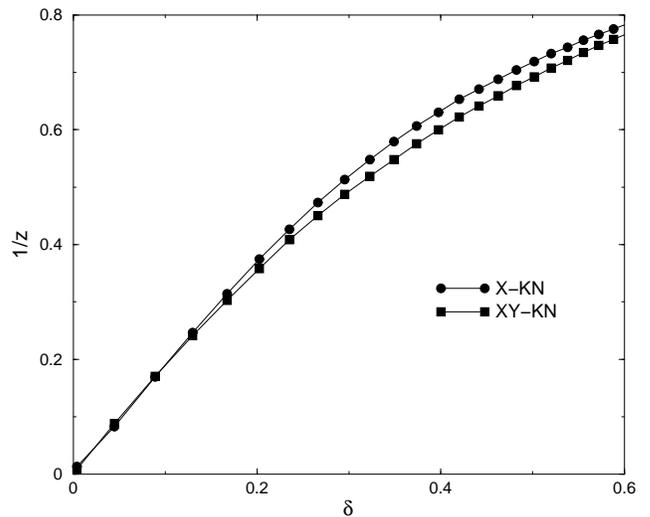} 
\caption{\label{fig3}The inverse of the 
 dynamic exponents $z$ for the $X-KN$ and $XY-KN$ models as function 
 of the distances to the random $QCP$ ($ \delta 
 >0$). They are obtained iterating numerically the recursion relations 
 starting from gapless rectangular distributions. The 
 dynamic exponents determine the singularities of the thermodynamic 
 quantities with decreasing temperature in the Griffiths phases (see 
 text).} 
\end{figure} 
 
Starting the iteration with  a distribution $P_J(J_i)$  with a gap, in the disordered Kondo 
phase ($J_0>1$) we obtain for both the $X-KN$ and $XY-KN$, besides the Griffiths 
phase for $\delta < \delta_G$, a strongly 
disordered {\em random Kondo phase} ($RKP$) for $\delta> \delta _{G}$. 
This phase consists essentially of a collection of isolated Kondo singlets with a distribution of excitation 
energies. It is natural then to describe such phase using a distribution of 
Kondo temperatures \cite{miranda}. 
 
In Fig.~4 we show  the uniform 
susceptibility as a function of temperature, $\chi (T)$, for the $X-KN$ 
model, calculated far away from the random $QCP$, deep in the disordered Kondo 
phase for two initial distributions $P_J(J_i)$. When the original distribution of Kondo couplings  
${J_{i}}$ has a finite gap $\Delta _{J}$ at 
low energies, the susceptibility (solid line) 
does not diverge at low $T$ as shown in  the figure.  
In this case the weakly disordered Griffiths phase extends to 
$\delta < \delta_G = 0.358$ which corresponds to a gap in the 
 original $P_J(J_i)$ of  $\Delta_J < 0.156$. 
The susceptibility shown is for $\delta = 0.44 > \delta_G$, in the $RKP$. It is clear that 
in this case the fixed point distribution of gaps or Kondo temperatures does 
not become sufficiently singular in the $RKP$ to yield a diverging 
susceptibility. 
In the same figure we show the temperature dependent susceptibility 
for an initial unbound distribution of interactions, i.e., 
a rectangular distribution with no gap.  The susceptibility now diverges as 
$T\rightarrow 0$. Since in this case of gapless distributions,  the Griffiths phase extends all over the 
disordered Kondo region, $\delta >0$,  the divergence of $\chi (T)$ is clearly 
associated with this Griffiths phase.  Similar behavior is found for the $XY-KN$ model. 
\begin{figure} 
 \includegraphics[width=7cm,angle=270]{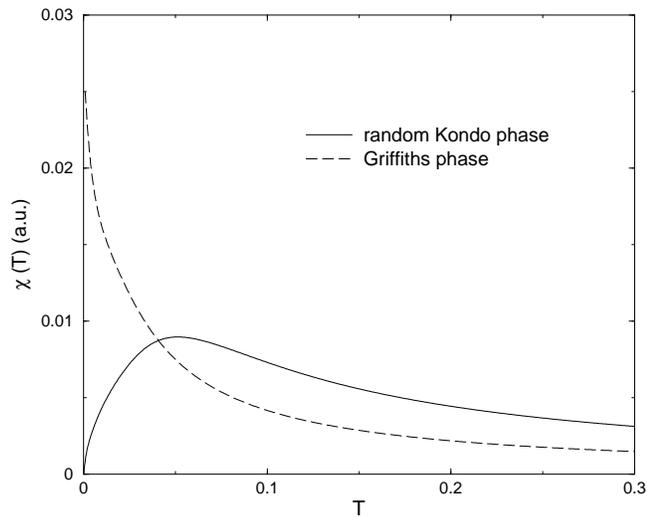} 
 \caption{\label{fig4}The uniform susceptibility of the $X-KN$ model as a 
function of temperature for two initial distributions of bonds 
and the same distance $\delta=0.44$ of the random $QCP$. For a 
gapped distribution  (solid line), for which $\delta  > \delta_G \approx 0.36$, and  
the system is in the $RKP$. 
For a gapless distribution  (dashed line) such that the system is in the 
Griffiths phase. Temperature is in units of $W_0$.} 
\end{figure} 
 
In summary, we have studied the effects of strong disorder in the $KN$ model 
for heavy fermion systems using a perturbative $RG$ approach. Differently from the case 
of weak disorder \cite{tati1}, strong disorder is a relevant perturbation 
and gives rise to Griffiths phases which for distributions with a gap 
extend over finite regions of the phase diagram, above and below the $QCP$. 
For $J_{0}>1$, in the strongly disordered regime above the Griffiths phase, 
we find a random Kondo phase which consists of a collection of isolated 
singlets with random excitation energies. It is natural to describe such a 
phase by a model with a distribution of Kondo temperatures \cite{miranda}. 
The susceptibility however does not 
diverge at low temperatures in the $RKP$. We observe a diverging susceptibility in the 
case of lower unbound distributions where the Griffiths phase extends all 
over the disordered region. Consequently our results indicate that a 
diverging $\chi(T)$ is due to a Griffiths phase. It can not be obtained in a 
model of isolated Kondo singlets with random Kondo temperatures.  
A fixed point distribution $P_J(J_i)$ of the power law type, that gives rise to
a diverging $\chi(T)$,   
is incompatible with the existence of isolated singlets.
In the Griffiths phases, the temperature dependence of the thermodynamic 
quantities is expressed in terms of the dynamic exponent $z$. We have obtained 
this exponent for both $KN$ models as a function of the distance to the $QCP$ 
for gapless distributions. 
 
At the random QCP of both the $X-KN$ and the $XY-KN$ 
there are random singlet phases characterized by thermodynamic quantities 
with power law behavior, arising from distributions of interactions like in 
Eqs. (\ref{potencia}) and (\ref{hopping}), with weakly temperature dependent exponents. For the 
disordered $X-KN$ model there is a disordered antiferromagnetic phase for $%
J_0 <1$ ($\delta <0$). The dense Kondo  phase for $J_0 >1$ ($\delta >0$) 
is similar in both $KN$ models 
studied. For the $XY-KN$ with $J_0<1$ ($\delta <0$) there is a random singlet phase 
coexisting for some range of $\delta$, which depends on the original distributions, 
with a weakly disordered Griffiths phase \cite{fisher2}. 
The exact mapping of the 
recursion relations for the $X-KN$ on those of the $RTIM$ allows to carry on 
the extensive results obtained on the latter to the present former model. 
It enables us to show unambiguously the existence of non-Fermi liquid behavior 
in strongly disordered heavy fermions associated with Griffiths phases. 
 
\begin{acknowledgments} 
The authors are grateful to the Brazilian agencies FAPERJ, CNPq and CAPES for 
financial support. 
\end{acknowledgments}

\end{document}